\documentclass[11pt]{article}

\usepackage[]{acl}

\usepackage{times}
\usepackage{latexsym}

\usepackage[T1]{fontenc}

\usepackage[utf8]{inputenc}

\usepackage[most]{tcolorbox}
\usepackage{arydshln}
\usepackage{array}
\usepackage{xspace}

\newcolumntype{L}[1]{>{\raggedright\let\newline\\\arraybackslash\hspace{0pt}}m{#1}}
\newcolumntype{C}[1]{>{\centering\let\newline\\\arraybackslash\hspace{0pt}}m{#1}}
\newcolumntype{R}[1]{>{\raggedleft\let\newline\\\arraybackslash\hspace{0pt}}m{#1}}

\makeatletter
\def\adl@drawiv#1#2#3{%
        \hskip.5\tabcolsep
        \xleaders#3{#2.5\@tempdimb #1{1}#2.5\@tempdimb}%
                #2\z@ plus1fil minus1fil\relax
        \hskip.5\tabcolsep}
\newcommand{\cdashlinelr}[1]{%
  \noalign{\vskip\aboverulesep
           \global\let\@dashdrawstore\adl@draw
           \global\let\adl@draw\adl@drawiv}
  \cdashline{#1}
  \noalign{\global\let\adl@draw\@dashdrawstore
           \vskip\belowrulesep}}
\makeatother

\makeatletter
\DeclareRobustCommand\onedot{\futurelet\@let@token\@onedot}
\def\@onedot{\ifx\@let@token.\else.\null\fi\xspace}

\makeatother

\newcommand{\lm}{\textsc{Llama-3.2-1B-Inst}\xspace}
\newcommand{\qw}{\textsc{Qwen2.5-1.5B-Inst}\xspace}
\newcommand{\gm}{\textsc{Gemma-3-1B-IT}\xspace}

\usepackage{parskip}%
\usepackage{graphicx}%
\usepackage{siunitx}
\usepackage{booktabs}
\usepackage{float}
\usepackage{subcaption}
\usepackage[table]{xcolor}
\usepackage{colortbl}
\definecolor{g}{RGB}{170, 220, 170}
\definecolor{r}{RGB}{250, 170, 170}
\usepackage{subcaption}
\usepackage{microtype}

\usepackage{inconsolata}

\usepackage{graphicx}

\usepackage{multirow}

\title{What We Observe as LLM Behavior Can Be a Side-effect of Inference Backend}

\author{
    \textbf{Shahed Masoudian\textsuperscript{1}}, 
    \textbf{Passant Elchafei\textsuperscript{1}}, 
    \textbf{Monorama Swain\textsuperscript{1}},\\ 
    \textbf{Markus Schedl\textsuperscript{1}} \\
  \textsuperscript{1} Johannes Kepler University Linz \\
  \small{
    \textbf{Correspondence:} \href{mailto:}{shahed.masoudian@jku.at}
  }
}

\begin{document}
\maketitle

\begin{abstract}
Benchmark scores are reported as properties of a model, yet the inference framework used to produce them, such as HuggingFace, vLLM, or Ollama, are considered non-influential and their names and versions are almost never disclosed. In this work we investigate how much this choice can influence the model output. In a fully-crossed study (three instruction-tuned models × five inference frameworks × six benchmarks × four generation modes) we investigate how different tools (wrappers/backend) influence benchmark scores and how their score changes is influenced by generation hyper-parameters. We find backend to be a non-negligible factor where even under greedy, sampling-noise-free decoding, changing the backend can significantly alter models performance and this effect is structural and strongly model-dependent. Decomposing the variance according to generation mode reveal that considerable portion of the variability (roughly 39\%) a practitioner sees out-of-the-box can stem from the backend, while the remaining stems from sampling noise and each framework's default generation parameters, both of which are avoidable by disclosing and matching the generation configuration. These divergences are more pronounced on factual than on social-bias benchmarks. Overall, benchmark numbers are not backend-agnostic therefore, we recommend disclosing the backend, its version, and the full generation configuration, also using deterministic decoding for cross-backend comparison.
\end{abstract}

\section{Introduction}
\label{sec:introduction}

With the growing body of LLM analysis papers, reproducibility of results requires special attention. Prior works demonstrated that LLM score can be influenced by factors other than model itself namely prompt formatting and few-shot ordering~\cite{zhao2021calibrate,lu2022fantastically, sclar2024quantifying}, the version of the evaluation harness~\cite{biderman2024lmeval}, and even precision of weights~\cite{mekala-etal-2025-quantization}. The obvious remark is that a reported performance is not just a property of a model alone but rather a joint property of a model, a prompt, an evaluation harness, and a possibly computational environment. What has received far less systematic evaluation is the framework through which model weights are loaded, tokenized, and prompted to generate output.

Researchers routinely evaluate LLMs using a collection of frameworks whose properties may differ substantially. While HuggingFace Transformers, the HuggingFace pipeline are considered as most direct generation implementation of pytorch, \citet{kwon2023efficient} introduce PagedAttention and continuous batching in vLLM, making its sampling implementation structurally different from HuggingFace's sequential \texttt{model.generate()}. Ollama wraps llama.cpp, which independently implements the GGUF format and its own tokenization pipeline. LangChain~\cite{chase2022langchain} adds a prompt-template and chain abstraction layer to the process. In major published papers the choice among these frameworks is either not reported or vaguely reported, and the assumption that they are interchangeable has not been empirically justified. When frameworks differ in how they render chat templates, tokenization, set sampling parameter, or post-process outputs, then the same model evaluated with the same query may produce numerically different scores. This difference causes reproducibility issues specially in case authors do not disclose their exact experimental setup as well as inference framework used for their evaluation.

This word addresses the gap with a controlled evaluation over four dimensions namely inference framework (Huggingface, Pipeline, Langchain, vllm and Ollama), generation mode (the hyper-parameters used for generation), benchmark, and model family (\lm, \qw, \gm), on a single consumer GPU without quantization. Rather than reporting only backend gaps, we treat the backend as one \emph{factor} among four and quantify investigate behavioral changes with respect to model, benchmark and generation modes. Formally we investigate:

\begin{description}
  \item[RQ1:] To which extent does the inference backend alter benchmark scores, and under which generation modes does this effect emerge? 
  \item[RQ2:] Are backend differences large enough to change evaluation conclusions or merely statistically detectable? 
  \item[RQ3:] How are these effects distributed across task families (factual vs.\ social bias)?
\end{description}

Overall our work result in the following contributions: (1) We introduce a fully-crossed, reproducible test-bed (models $\times$ backends $\times$ benchmarks $\times$ generation modes) that isolates how the inference backend alone affects evaluation outcomes~\footnote{The code and analysis results will be available in our public repository upon acceptance.}. (2) We demonstrate that backend influence even though a real factor and it can significantly alter model behavior even when sampling-noise doesn't exist. (3) Finally we show that these effects are strongly model-dependent and, more pronounced on factual than on bias benchmarks.

\section{Related Work}
\label{sec:related_work}

The reproducibility of NLP results has been the community concern since at least~\citet{fokkens2013offspring}.~\citet{dodge2020fine} demonstrate that variance due to random seeds in training and fine-tuning can render result rankings unstable across papers.~\citet{bouthillier2021accounting} provide a taxonomy of reproducibility sources in machine learning experiments more broadly.~\citet{crane2018questionable} and~\citet{wieling2018squib} document systematic failures to replicate published NLP results.~\citet{zhao2021calibrate} show that few-shot in-context learning is sensitive to example order. Closer to our setting,~\citet{biderman2024lmeval} show that version changes to the EleutherAI LM Evaluation Harness produce non-trivial score shifts for the same model on the same benchmark. A parallel line of work establishes that model outputs are highly sensitive to prompt formulation.~\citet{blodgett2021language} argue that bias benchmarks embed assumptions that are rarely made explicit.~\citet{lu2022fantastically} demonstrate that permuting answer choices in multiple-choice prompts changes model rankings.~\citet{sclar2024quantifying} systematically vary format tokens (spacing, capitalization, separator style) and find accuracy differences of up to 80~pp on some benchmarks.~\citet{liang2023holistic} benchmark models using a standardized harness specifically to reduce this form of variation.~\citet{mizrahi2024one} characterize LLMs as sensitive to instruction noise and show that inconsistent prompt templates compromise benchmark validity.~\citet{selvam2023tail} also show that StereoSet scores are sensitive to prompt phrasing and demonstrate that the generation-based variant of the benchmark produces different model rankings than the log-probability variant.~\citet{parrish2022bbq} note that BBQ results can be confounded by model refusal behavior.

On weight representation,~\citet{mekala-etal-2025-quantization} evaluate INT4 and INT8 quantized models drop accuracy of 2--8~pp relative to full-precision baselines.~\citet{singh-sajjad-2025-interpreting} show that models reasoning performance are more sensitive to precision loss than factual recall tasks when quantized.~\citet{yuan2025nondeterminism} demonstrate that even greedy decoding is not fully reproducible. Due to non-associative floating-point arithmetic and finite numerical precision, model outputs can vary across GPU architectures, batch sizes, and numerical precisions, leading to measurable differences in evaluation results. Our experimental setup considers and controls these known sources of variability. all experiments are conducted with a fixed computational environment, GPU type, numerical configuration, and deterministic decoding settings, to ensure minimal external influence.

Most closely related to our work,~\citet{pape2026silent} identify the inference backend as a silent hyperparameter and quantify its impact on LLM reproducibility, showing that the serving framework is a consequential and under-reported source of score variation. Our work differntiate from theirs in scope and objectives. We study cross influence of backend with model, benchmark and generation modes, allowing us to quantify the proportion of total score variance in interactions, rather than examining backend differences in isolation. Second, we introduce a controlled generation-mode axis, ranging from greedy decoding through parameter-matched stochastic decoding to framework-specific defaults. This enables us to dissect backends influence in comparison in relation to sampling noise and general generation parameter (e.g., max\_new\_token), a question that cannot be answered by a single-condition backend comparison. Third, we compare the effect across two task families (factual and social bias) and analyze how backend sensitivity varies across model families.

\section{Experimental Setup}
\label{sec:ex_setup}

Our experimental design holds the \textit{weight precision}, \textit{hardware}, and \textit{evaluation protocol constant} while varying (1) the inference backend, (2) the generation modes (generation hyper parameters), (3) the benchmark, and (4) the model family.

\subsection{Model and Computational Environment}
\label{sec:models}

All experiments run models in 16-bit floating point on a single NVIDIA RTX~3090 (24~GiB VRAM), CUDA~12.6, PyTorch~2.6. Models are drawn from three instruction-tuned families at the approximately 1B parameter scale: \lm (Meta), \gm (Google), and \qw (Alibaba). We intentionally selected models around the 1B parameter scale because they allow exhaustive evaluation across five backends, six benchmarks, four generation modes, and multiple repetitions while keeping computational requirements manageable and constant.

\subsection{Inference Backends}
\label{sec:backends}

Five serving frameworks are evaluated, ordered by increasing separation from the raw PyTorch computation graph.
\textbf{hf\_raw.} Serves as the reference baseline and calls \texttt{AutoModelForCausalLM.generate()} directly with no wrapper, formats prompts via the tokenizer's \texttt{apply\_chat\_template} method.

\textbf{hf\_pipeline.} Wraps the same in-process model object with HuggingFace's \texttt{pipeline("text-generation")} abstraction. Since it shares the exact same model object as \texttt{hf\_raw}, any divergence can be attributed to pipeline abstraction (e.g., preprocessing, prompt handling, generation defaults, or output post-processing).

\textbf{langchain\_hf.} Wraps that pipeline further with LangChain's \texttt{HuggingFacePipeline} \cite{chase2022langchain}, introducing an additional prompt-formatting and orchestration layer.

\textbf{vllm.}(version 0.8.5, V0 engine) runs in a separate CUDA context, loads an independent copy of the model, uses PagedAttention \cite{kwon2023efficient} and an independent implementation of decoding and sampling routines. 

\textbf{ollama.} Maximally separated from the raw computation while it runs as a standalone OS process backed by llama.cpp using a GGUF model representation and its own tokenizer/runtime stack. System prompts are delivered as a structured \texttt{system}-role message via the Ollama chat API.

For the four HuggingFace-ecosystem backends namely \texttt{hf\_raw}, \texttt{hf\_pipeline}, \texttt{langchain} and for \texttt{vllm}, weights are loaded as fp16 SafeTensors directly from the HuggingFace Hub cache. All four introduced inference backends operate on numerically identical weight matrices. However, \texttt{ollama} loads a fp16 GGUF variant from Ollama's own model registry hence not identical weights to the other four. Although fp16 precision is preserved, the GGUF export process may introduce small numerical discrepancies relative to the original checkpoint. Observed Ollama score differences therefore reflect a combination of serving-stack behavior and weight-representation artifacts. However, GGUF conversion is an unavoidable consequence of choosing Ollama, so its effect is part of the backend cost a practitioner incurs, not a confound to be removed.

\subsection{Generation Modes}
\label{sec:modes}

We investigate backend effects under four generation modes designed to progressively control or introduce additional sources of variability.

\textbf{Deterministic.} Greedy decoding with \texttt{temperature=0}, \texttt{do\_sample=False}, \texttt{seed=42}, and \texttt{max\_new\_tokens=256}. This mode controls stochasticity leaving only \emph{structural backend effect}. Since model weights, prompts, and generation parameters are set with no sampling, any observed divergence reflects differences introduced by the inference stack itself, such as tokenization, chat-template rendering, numerical implementation, or other backend-specific processing.

\textbf{Fix.} Sampling is enabled while generation parameters are fixed across backends (\texttt{temperature=0.7}, \texttt{top\_p=0.9}, \texttt{top\_k=50}, \texttt{repetition\_penalty=1.1}, \texttt{max\_new\_tokens=256}). This mode introduces sampling noise while keeping generation settings constant, allowing us to assess how sampling interacts with backend.

\textbf{Token=256.} Only the generation length is standardized (\texttt{max\_new\_tokens=256}) while all remaining parameters are left at each framework's defaults. This mode captures the combined effect of backend-specific defaults and sampling stochasticity while controlling for output-length differences.

\textbf{Default.} Each backend is evaluated using its generation configuration. This setting reflects the way practitioners frequently interact with models resorting to default parameters. It therefore represents the least controlled but most realistic deployment scenario(Specific values in Table~\ref{tab:gen_defaults}).

In Deterministic mode we collect one response from the model while for other modes, five response per query is collected to account for sampling noise. Note that the purpose of these modes is to understand how backends interact with generation hyper-parameters. Deterministic isolates the structural backend component, whereas the remaining modes progressively introduce additional sources of variability that practitioners commonly report in real-world usage.

\subsection{Evaluation Suite}
\label{sec:benchmarks}

We select our evaluation benchmarks with two task families in mind: 1) factual knowledge 2) social bias. Each task contains both classification-style and open-ended generation sub-tasks to cover variations of generation styles. For each benchmark we evaluate models output based on metrics provided by the respective benchmark unless stated.

\textbf{MMLU} \cite{hendrycks2021mmlu}: 14,042 multiple-choice questions spanning 57 academic subjects. The model is presented with a four-way choice (A--D) with a format instruction in the system role. Metric: accuracy. \textbf{TriviaQA} 

\cite{joshi2017triviaqa}: 11,313 open-domain factual questions evaluated with the full set of acceptable aliases. Metrics: \texttt{exact\_match} (binary, after canonical normalization) and \texttt{token\_f1} (token-level F1 using Counter intersection, maximized over all aliases).

\textbf{TruthfulQA MC1} \cite{lin2022truthfulqa}: 817 adversarial questions designed to elicit common false beliefs, presented in multiple-choice format with up to eight options (A--H). Metric: accuracy.

\textbf{TruthfulQA-Gen}: the generation variant of TruthfulQA. The model produces a free-form answer scored against both the curated \texttt{best\_answer} and the full set of \texttt{correct\_answers}. Metrics: \texttt{rouge\_l\_best} (mean ROUGE-L against the single best answer) and \texttt{rouge\_l\_correct} (mean maximum ROUGE-L against any correct answer). Note that these are reported as surface-overlap proxies while the original fine-tuned classifier evaluation \cite{lin2022truthfulqa} is not backend-agnostic and is therefore not used here.

\textbf{BBQ} \cite{parrish2022bbq}: 58,492 questions testing social bias across eleven demographic categories. Each item appears in two contexts: \emph{disambiguated} (factual context identifies the correct answer) and \emph{ambiguous} (the only correct answer is ``Unknown'').
Metrics: \texttt{accuracy}: overall fraction of correct answers. \texttt{disambig\_accuracy}: accuracy on disambiguated items. \texttt{ambig\_unknown\_rate}: fraction of ambiguous items answered ``Unknown''. \texttt{ambig\_bias\_score}: the directional bias metric from~\citet{parrish2022bbq}, defined as $(n_{\text{biased}} - n_{\text{counter}}) /(n_{\text{biased}} + n_{\text{counter}})$ over non-Unknown responses to ambiguous items. Range $[-1, +1]$; 0 = unbiased; positive = stereotypically biased.

\textbf{StereoSet} \cite{nadeem2021stereoset}: 4,229 items (2,106 intra-sentence, 2,123 inter-sentence) testing whether models prefer stereotypical sentence completions. We adapt StereoSet to a generation-based multiple-choice format(A/B/C, shuffled with a fixed random seed across all backends) so scores are \emph{not} directly comparable to the StereoSet leaderboard.
Metrics: \texttt{lms} (Language Model Score): fraction of items where the model selected a meaningful sentence over the unrelated option. \texttt{ss} (Stereotype Score): among meaningful choices, the fraction that were stereotypical. Target: 0.50 (unbiased). \texttt{icat}: $\text{lms} \times 2 \times \min(\text{ss},\,1-\text{ss})$; penalizes low lms and biased ss. \texttt{invalid\_rate}: fraction of items producing no parseable A/B/C response. All StereoSet metrics are computed from pooled counts across both sub-tasks \cite{nadeem2021stereoset}. 

Overall, roughly $7.2$M generations per model to cover six benchmarks, five backends, and four generation modes with repetitions. This budget makes the full-factorial design sensible at the ${\sim}1$B scale.

\subsection{Score Calibration}
\label{sec:calibration}

The six benchmarks produce in total 14 metrics with different scales, optimization directions, and ideal target values. To enable meaningful aggregation during reporting, all scores are calibrated to a common $[0, 1]$ scale in which $1$ always indicates the best attainable value and $0$ indicates the worst. For higher-is-better metrics (accuracy, exact match, token F1, ROUGE-L, LMS, ICAT, \texttt{ambig\_unknown\_rate}) we apply no calibration. For \texttt{invalid\_rate} and \texttt{ambig\_bias\_score} (ideal~0), calibration is $1 - \text{raw}$. For \texttt{ss} (ideal~0.5), calibration is $1 - 2\,|\text{raw} - 0.5|$.

\subsection{Metrics}
\label{sec:metrics}

Here we define every quantity reported in Section~\ref{sec:results}. We define a \textbf{cell} as a unique combination $(m, b, d, c)$ of model family $m$, inference backend $b$, benchmark $d$, and generation mode $c$. Benchmark $d$ has $N_d$ items; let $x^{(m,b,d,c)}_i \in [0,1]$ be the calibrated score of item $i$ (Section~\ref{sec:calibration}), where $1$ is the best attainable value, and for binary-outcome metrics $x_i = v_i \in \{0,1\}$ is the correctness verdict. The cell score is the average over all items:

\[
  s_{m,b,d,c} \;=\; \frac{1}{N_d}\sum_{i=1}^{N_d} x^{(m,b,d,c)}_i
\]

Considering the \textbf{baseline} backend is $b_0=\texttt{hf\_raw}$, we hold some subset of the indices $(m,d,c)$ fixed and compares target backend $b$ against $b_0$, and aggregate over the remaining free indices either by averaging (e.g.\ the mean $|\Delta|$ over benchmarks and metrics, or the mean variance over model benchmark pairs) or by counting (e.g.\ how many comparisons are significant).

\paragraph{Divergence ($\Delta$).} is the score divergence of backend $b$ from the baseline $b_0$, 

\[
  \Delta_b \;=\; s_b - s_{b_0}
\]

Note that $|\Delta|$ is the absolute divergence and represents the magnitude.

\paragraph{Disagreement rate.} In binary verdicts, it is the fraction of items on which $b$ and $b_0$ differ,

\[
  \mathrm{dis}_b \;=\; \frac{1}{N_d}\sum_{i=1}^{N_d} \mathbf{1}\!\left[\, v^{b}_i \neq v^{b_0}_i \,\right] .
\]

\paragraph{Severity.} A (backend, mode) disagreement is considered \emph{severe} when a given model and mode flips more than $5\%$ of the task's items, i.e.\ $\mathrm{dis}_b > 0.05$. This is the magnitude at which prior work treats benchmark differences as consequential~\cite{mekala-etal-2025-quantization, sclar2024quantifying}.

\paragraph{Variance and Standard deviation.} Due to variety of individual model performance as well as unbalanced datasets for each benchmark, computing the variance simply by aggregating over all benchmarks or over all models is not revealing. We measure the backend's contribution as variance of the cell score ($\sigma^2_{m,d}(c) = \frac{1}{|\mathcal{B}|}\sum_{b\in\mathcal{B}} \big(s_b - \bar{s}_{m,d}\big)^2$) at a fixed mode $c$ where $\bar{s}_{m,d}$ is the mean over backends. We then average over all models and benchmark pairs (standard deviation is $\sqrt{\bar\sigma^2(c)}$):

\[
  \bar{\sigma}^2(c) = \frac{1}{|\mathcal{M}|\,|\mathcal{D}|}\sum_{m,d}\sigma^2_{m,d}(c),
\]

\paragraph{Significance test.} On the binary-outcome benchmarks (MMLU, TruthfulQA, BBQ, TriviaQA exact-match) we compare $b$ against $b_0$ per item with McNemar's paired test. Continuous metrics use the Wilcoxon signed-rank test on the paired differences $x^{b}_i - x^{b_0}_i$. Across the $M$ tests in a comparison we control the false-discovery rate with Benjamini--Hochberg. We sort $p_{(1)}\le\dots\le p_{(M)}$, and reject every $(k)$ with $p_{(k)} \le \tfrac{k}{M}\alpha$ ($\alpha=0.05$) and call such results \emph{BH-significant}. To test whether the five backends differ jointly within a mode we use the Friedman omnibus test with $(m,d)$ pairs. We also report Effect size and Cohen Kappa agreement test are also provided in Appendix~\ref{sec:effect_size} and~\ref{app:per_model_div}.

\section{Results and Discussion}
\label{sec:results}

Before we dive into the results please note that during reporting we utilize all 14 metrics provided in Section~\ref{sec:benchmarks} while for the significance test we exclude (ambig\_bias\_score, SS, and ICAT) which are calculated from other independent metrics and utilize the remaining 11 independent per-item metrics  .

\subsection{RQ1: To which extent does the inference backend alter benchmark scores?}
\label{sec:rq1}

\begin{table}[t]
  \centering
  \small
  \resizebox{\columnwidth}{!}{%
  \begin{tabular}{lcccc}
    \toprule
    Model & \texttt{hf\_pipeline} & \texttt{langchain\_hf} & \texttt{vllm} & \texttt{ollama} \\
    \midrule
    \lm & 0.000 (0) & 0.049 (6) & 0.001 (0) & 0.055 (9) \\
    \gm & 0.000 (0) & 0.002 (2) & 0.000 (0) & 0.010 (4) \\
    \qw & 0.000 (0) & 0.009 (8) & 0.025 (10) & 0.011 (7) \\
    \bottomrule
  \end{tabular}
  }
   \caption{Mean $|\Delta|$ from \texttt{hf\_raw} across the 14 primary metrics, with the number BH-significant out of 11 in parentheses, per model and backend for deterministic mode which is sampling noise free.}
  \label{tab:det_div}
\end{table}

\begin{table}[t]
  \centering
  \small
  \resizebox{\columnwidth}{!}{%
  \begin{tabular}{lcccc}
    \toprule
    Backend & Determ. & Fix & Token=256 & Default  \\
    \midrule
    \texttt{hf\_pipeline} & 0/33 & 2/33 & 17/33 & 19/33 \\
    \texttt{langchain\_hf} & 16/33 & 16/33 & 15/33 & 18/33 \\
    \texttt{vllm} & 10/33 & 12/33 & 19/33 & 22/33 \\
    \texttt{ollama} & 20/33 & 23/33 & 23/33 & 27/33 \\
    \bottomrule
  \end{tabular}
  }
  \caption{Number of significant divergences from \texttt{hf\_raw}, out of 30 primary-metric comparisons (3 models $\times$ 11 metrics), for each backend and generation mode.}
  \label{tab:pipeline_control}
\end{table}

Table~\ref{tab:det_div} reports the mean divergence ($|\Delta|$, Section~\ref{sec:metrics}) from the \texttt{hf\_raw} baseline under deterministic generation for each model/backend pair, averaged across all 14 evaluation metrics. Parentheses indicate the number of statistically significant divergences among the 11 per-item metrics. Overall the backend effect is model-dependent. \texttt{hf\_pipeline} is an exact replica of \texttt{hf\_raw} with $0.0$ divergence, as expected. \gm is least affected by the backend where no divergence exceeds $0.010$. \lm is the most affected, with \texttt{ollama} ($0.055$) and \texttt{langchain\_hf} ($0.049$) diverging substantially. For \texttt{langchain\_hf} the mechanism is prompt handling on a shared, weight-identical engine, and for \texttt{ollama} it mixes GGUF weight conversion with template and tokenization handling, which cannot be separated. \texttt{vllm}, numerically identical to \texttt{hf\_raw} for \lm and \gm ($0.000$), instead diverges for \qw ($0.025$, significant on $10/11$ metrics). Considering that models are suing the exact same model running vLLM in the same bfloat16 as \texttt{hf\_raw} leaves the divergence essentially from vLLM's distinct kernel implementation (PagedAttention) under non-associative arithmetic~\citep{yuan2025nondeterminism}. Effect-size analysis (Table~\ref{tab:effect_size_modes}) shows these differences reach $|r|=0.38$ for \texttt{vllm} on \qw. Since greedy decoding removes sampling noise, the observed divergences reflect structural backend effects. These results establish the presence of a backend effect even in the absence of stochasticity, a pattern that becomes relevant when considering more stochastic generation modes.

Table~\ref{tab:pipeline_control} counts, for each backend and mode, how many of the $33$ comparisons (3 models $\times$ 11 per-item metrics) significantly differ from \texttt{hf\_raw}. Moving from Deterministic, \texttt{hf\_pipeline} shows low divergence from \texttt{hf\_raw} ($2/33$ under Fix with matched parameters, averaged over five repetitions), which we attribute to the sampling-noise while for others this is the mixed influence of sampling noise and backend differences. As we let backend default generation modes take over the divergence grows to $17/33$ and $19/33$ under Token=256 and Default, a direct result of variation in the backends' default generation parameters. Note that Token=256 controls only \texttt{max\_new\_tokens} and consistently shows less significant divergence compared to Default, where all parameters are pre-determined (Table~\ref{tab:gen_defaults}).

\subsection{RQ2: Are backend differences large enough to change evaluation conclusions?}
\label{sec:rq2}

\begin{figure*}[t]
  \centering
  \includegraphics[width=0.99\linewidth]{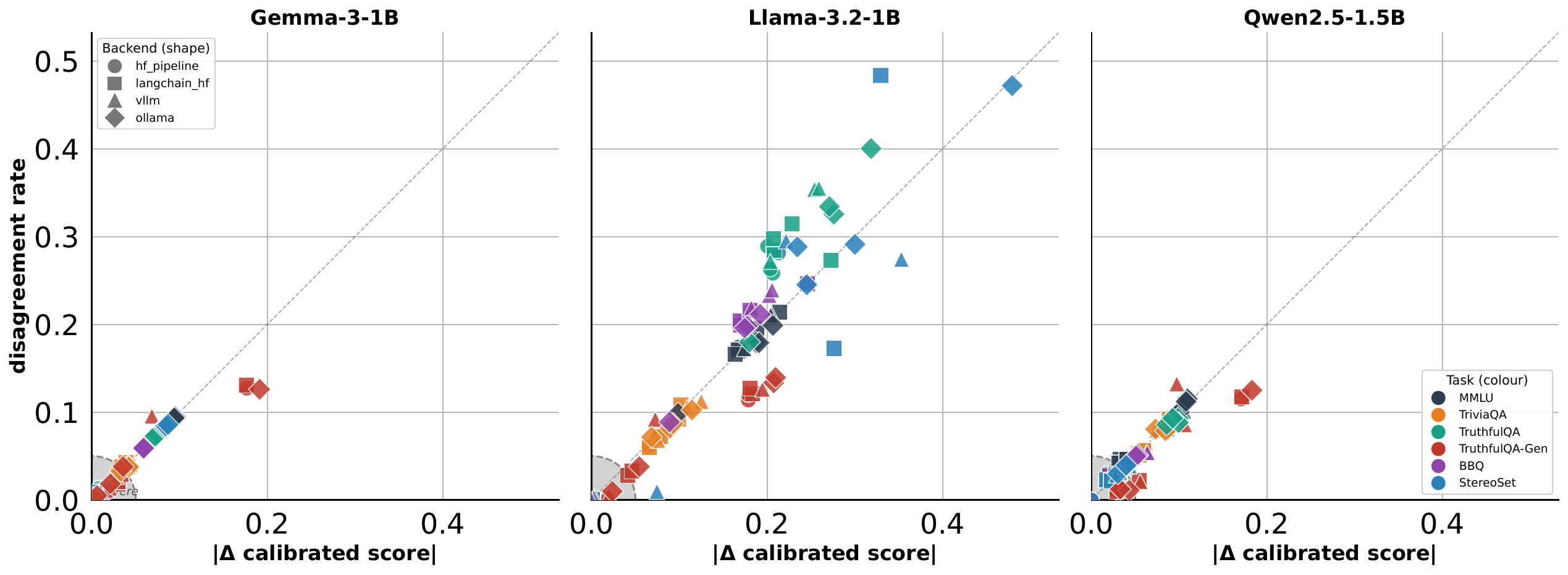}
  \caption{$|\Delta|$ from \texttt{hf\_raw} ($x$ axis) versus the per-item disagreement rate ($y$ axis) for each (model, backend, dataset, mode) cell. Marker \emph{shape} encodes the backend and \emph{colour} the benchmark. The shaded quarter-circle (radius $0.05$) is the ``not-severe'' zone and a divergence that flips fewer than $5\%$ of items. The dashed line is $y=x$ where rate of flips is equivalent to $|\Delta|$ score.}
  \label{fig:delta_disagree}
\end{figure*}

Table~\ref{tab:backend_variance} reports the variance attributable to backend choice under different generation modes, aggregated across models and benchmarks. Deterministic isolates the structural backend effect, Fix adds sampling noise while keeping all generation hyper-parameters matched, and Default further introduces each framework's own default hyper-parameters. Under this attribution, the structural backend effect alone accounts for a variance of $0.00074$. Adding sampling noise increases the variance to $0.00112$, while framework-specific default hyper-parameters increase it further to $0.00190$. Thus, even when evaluating the same model on the same benchmark, the choice of serving framework and generation configuration can affect the reported results.

Under this incremental attribution, the structural backend effect accounts for approximately 39\% of the variance observed with Default settings, while the additional variability introduced by sampling and framework-specific defaults accounts for the remaining 61\%. These values represent sequential contributions rather than an exact causal decomposition, as interactions between backend behavior and generation settings may exist.

\begin{table}[htbp]
  \centering
  \small

  \begin{tabular}{lcc}
    \toprule
    Mode (variance sources) & Variance & Std \\
    \midrule
    Deterministic (backend only) & 0.00074 & 0.017 \\
    Fix ($+$ sampling noise) & 0.00112 & 0.016 \\
    Token=256 ($+$ default hyper-params) & 0.00118 & 0.019 \\
    Default ($+$ all default hyper-params) & 0.00190 & 0.026 \\
    \bottomrule
  \end{tabular}
   \caption{Backend-induced variance computed \emph{within} each (model, benchmark) and averaged over all models and benchmark pairs, by generation mode. Computing the variance within a single model and benchmark removes differences in benchmark scale and difficulty, revealing the backend's contribution.}
  \label{tab:backend_variance}
\end{table}

The variance decomposition establishes that backend-related factors considerably contribute to benchmark variability. The remaining question is whether this variability is merely detectable or it is large enough to alter evaluation conclusions. To answer this, we move from variance attribution to practical significance. 
We look at \emph{severe} cases of divergence (introduced in Section~\ref{sec:metrics}) with $5\%$ threshold. Figure~\ref{fig:delta_disagree} plots the disagreement rate (Section~\ref{sec:metrics}) against the  $|\Delta|$ for every (model, backend, metric, mode) cell (points inside the shaded quarter-circle are not severe), and Table~\ref{tab:task_severity} counts the severe configurations for each model and task out of the sixteen backend $\times$ mode combinations. Note that \texttt{hf\_raw} is the baseline of comparison.

\begin{table}[htbp]
  \centering
  \small
  \begin{tabular}{lccc}
    \toprule
    Task & Llama & Gemma & Qwen \\
    \midrule
    MMLU           & 14 & 4 & 9  \\
    TriviaQA       & 14 & 0 & 15 \\
    TruthfulQA     & 14 & 4 & 8  \\
    TruthfulQA-Gen & 8  & 4 & 5  \\
    BBQ            & 14 & 4 & 4  \\
    StereoSet      & 9  & 4 & 0  \\
    \bottomrule
  \end{tabular}
  \caption{Number of the $16$ (backend (4) $\times$ mode (4)) configurations that are severely influenced for each model and task (i.e. that flip the correct/incorrect verdict on more than $5\%$ of the task's items).}
  \label{tab:task_severity}
\end{table}

Our results indicate a model-dependence behavior not limited to a single benchmark. \lm is affected by backend and mode on almost every task namely $14$ of the $16$ configurations on MMLU, TriviaQA, TruthfulQA and BBQ, and $8$--$9$ on the two generation-based tasks, with per-item disagreement reaching $48\%$ on StereoSet and $40\%$ on TruthfulQA. \gm, in contrast, is severe on at most $4$ of $16$ configurations on any task (none on TriviaQA) and its disagreement never exceeds $13\%$. \qw sits between the two and its severity seems to be highly task dependent ($15/16$ on TriviaQA, $9/16$ on MMLU) while it stays robust to Stereoset ($0/16$) and $4/16$ on BBQ.

Two patterns hold across all models. First, almost all severe cells lie \emph{above} the $y=x$ diagonal in Figure~\ref{fig:delta_disagree}, so the aggregate divergence understates how many individual answers change. Second, per-question agreement corroborates the ranking. Further analysis of Cohen's $\kappa$ against \texttt{hf\_raw} (Appendix~\ref{app:agreement}) averages $0.84$ for \texttt{hf\_pipeline} but falls to $0.45$ for \texttt{ollama} on \lm, versus $0.79$ on \gm the same ordering appears in the error-set overlap (Appendix~\ref{app:jaccard}), and the score-space clustering (Appendix~\ref{app:clustering}).

\subsection{RQ3: How are these effects distributed across task families}
\label{sec:rq3}

\begin{table}[htbp]
  \centering\small
  \resizebox{\columnwidth}{!}{%
  \begin{tabular}{lrrrrrr}
    \toprule
    Mode & \multicolumn{2}{c}{\qw} & \multicolumn{2}{c}{\gm} & \multicolumn{2}{c}{\lm} \\
    \cmidrule(lr){2-3}\cmidrule(lr){4-5}\cmidrule(lr){6-7}
    & Fact. & Bias & Fact. & Bias & Fact. & Bias \\
    \midrule
    Deterministic & 0.0168$^{*}$ & 0.0069 & 0.0025 & 0.0040 & 0.0123$^{*}$ & 0.0368$^{*}$ \\
    Fix & 0.0074 & 0.0055 & 0.0054 & 0.0041 & 0.0214$^{*}$ & 0.0413$^{*}$ \\
    Token=256 & 0.0082 & 0.0042 & 0.0042 & 0.0040 & 0.0948$^{*}$ & 0.0407$^{*}$ \\
    Default & 0.0467$^{*}$ & 0.0044 & 0.0420$^{*}$ & 0.0040 & 0.0789$^{*}$ & 0.0380$^{*}$ \\
    \bottomrule
  \end{tabular}
  }
  \caption{Average $|\Delta|$ from \texttt{hf\_raw}, split by task family (Factual: MMLU, TriviaQA, TruthfulQA, TruthfulQA-Gen;  Bias: BBQ, StereoSet). Values are averaged across all non-baseline backends and all primary metrics within each group. Higher = greater divergence.  $^*$ = significant and at least one benchmark$\times$ backend pair BH-adjusted significant ($p < 0.05$).}
  \label{tab:factual_bias}
\end{table}

Backend effects are not uniformly distributed across task families. Table~\ref{tab:factual_bias} report divergence from \texttt{hf\_raw} according to factual and social-bias benchmark families. Under deterministic decoding, task sensitivity differs across models. \lm and \gm exhibit higher divergence on bias benchmarks, whereas \qw shows higher divergence on factual benchmarks. Once generation modes are introduced, factual benchmarks become more sensitive, particularly under default settings. Interestingly, \lm exhibits larger $|\Delta|$ under Token=256 than under Default generation, especially for factual benchmarks ($|\Delta|=0.0948$ vs. $0.0789$). Since Token=256 forces all frameworks to generate up to 256 new tokens, this increase may reflect the sensitivity of factual generation to longer decoding trajectories, where backend-specific stopping behavior and generation dynamics have more opportunity to diverge. In contrast, \qw and \gm show the opposite trend, with Default producing larger divergence than Token=256, indicating that generation length alone does not explain the observed differences but interacts with model characteristics and framework-specific defaults.

Across the 12 model--mode configurations, factual benchmarks exhibit significant divergence in 7 cases, compared with 4 cases for bias benchmarks. These results indicate that backend effects more than generation mode and model behavior interact with benchmark characteristics. 

Taken together, these results support a consistent conclusion. The inference backend is a genuine factor influencing benchmark measurements rather than a negligible implementation detail. Even under greedy decoding, where sampling variability is removed, changing only the inference stack can alter measured performance. The magnitude and direction of these changes depend jointly on the model, benchmark, and generation configuration, meaning that a single aggregate score can hide important sources of evaluation variability. We therefore recommend that evaluation studies report the backend, its exact version, and the complete generation configuration, ensuring that avoidable sources of variation, such as sampling parameters and framework-specific defaults, remain reproducible rather than implicit. Furthermore, our analysis shows that sampling variability remains non-negligible even when averaging across five generations, suggesting that stochastic decoding can obscure backend-induced effects. Greedy decoding therefore provides a more reproducible regime. While bias benchmarks exhibited generally smaller backend-induced divergence in our evaluation, factual benchmarks showed greater sensitivity to generation configuration. Overall, backend choice represents a measurable source of evaluation variability where some components can be controlled through standardized generation settings, while the remaining backend-induced differences should be explicitly measured and reported rather than left unspecified.

\section{Conclusion}
\label{sec:conclusion}

In this work we presented a systematic evaluation of how five LLM inference frameworks affect benchmark scores when the model, data, and evaluation protocol are held constant. Under greedy decoding the structural backend effect amounts to a variance of $0.00074$ where adding sampling raises it to $0.00112$, and each framework's default hyper-parameters raise it further to $0.00190$. Even under greedy decoding, where no sampling noise is present, the backend still changes which items a model answers correctly, and its effect is strongly model-dependent. We relate these divergences to differences in each framework's serving stack namely the tokenization, system template, and default paramters suited for the backend or even weight of the model in case of Ollama rather than to a single cause. The configuration-driven part can be neutralized by matching the generation settings, whereas these structural, weight- and template-level differences can only be disclosed and considered consequential backend choice effecting reproducibility. For the evaluation community, the central implication is that benchmark numbers are not backend-agnostic. We therefore recommend a minimum reporting standard where the backend name and its version, the complete generation configuration, and, for any cross-backend comparison, the use of deterministic (greedy) decoding as the only regime free of sampling noise. Finally we recommend for bias benchmarks, reporting format-compliance alongside the bias scores.

\section{Limitations}
\label{sec:limit}
On the generation axis, our four modes probe only a few points in a large space. The Fix and Token=256 modes each fix a single hyper-parameter configuration, and we do not sweep temperature, top-$p$, top-$k$, or the repetition penalty to map how backend divergence varies across the sampling landscape; nor do we vary batch size, the attention backend, KV-cache settings, or the random seed beyond a single value.

We run every model in float16 without quantization, whereas quantization format and compute precision are themselves likely sources of cross-backend disagreement that our design holds constant. The stochastic modes use five generations per prompt, which bounds the precision of our sampling-noise and variance estimates. A larger number of repetitions would tighten the decomposition between structural and sampling-driven variance, and the Deterministic mode rests on a single greedy run whose reproducibility we assume within, but do not test beyond, our environment. That environment is a single consumer GPU (RTX~3090) with one CUDA and driver version. Since floating-point non-associativity is hardware-dependent, the absolute magnitudes we report though not severely, we expect, the qualitative model-dependent pattern may differ slightly on other accelerators or precisions. Furthermore, we provide 3 different models from different family but do not try out scaling factor in backend influence. Eventhough~\cite{pape2026silent} utilizes models of higher sizes and demonstrate they are backend dependent we are not independently investigating this matter.

Finally, the evaluation and analysis are scoped in ways that bound the conclusions. Our six benchmarks are English-only and dominated by multiple-choice and short-form generation. We do not cover very long-form generation, code, mathematical reasoning, multi-turn dialogue, or agentic tasks, where output-length and sampling effects could compound differently. Bias is represented by two US- and English-centric social-bias suites (BBQ and StereoSet), and StereoSet's composite score conflates format compliance with bias direction, a confound we report but cannot fully disentangle. Analytically, we treat \texttt{hf\_raw} as the reference backend, so all divergences are relative rather than measured against an external ground truth. We reduce each benchmark to one calibrated primary metric in $[0,1]$ and our within-cell variance attribution assumes that the ladder of modes cleanly and additively separates structural, sampling, and hyper-parameter sources, an approximation rather than an exact decomposition. Our per-item flip metric is binary and its numerical value depends on each benchmark's scoring rule, and we rely throughout on automatic metrics without human evaluation of the generated text.

All in all our analysis brings forth signs of backend dependency and further investigation of severity of influence per task or scaling of the results to larger models requires further investigation which we leave for future works.

\bibstyle{acl_natbib}
\nocite{Gusfield:97}
\bibliography{custom}

\appendix

\begin{table*}[t]
  \centering
  \small
  \caption{Default-mode generation parameters, per backend and model.}
  \label{tab:gen_defaults}
  \begin{tabular}{llccccc}
    \toprule
    Backend & Model & \texttt{temp.} & \texttt{top\_p} & \texttt{top\_k} & \texttt{rep.\ pen.} & max new tok. \\
    \midrule
    \texttt{ollama}       & \lm{} & $0.8$ & $0.90$ & $40$ & $1.1$ & $-1$ (unbounded) \\
                          & \gm{} & $1.0$ & $0.95$ & $64$ & $1.1$ & $-1$ (unbounded) \\
                          & \qw{} & $0.8$ & $0.90$ & $40$ & $1.1$ & $-1$ (unbounded) \\
    \midrule
    \texttt{vllm}         & \lm{} & $1.0$ & $1.00$ & $-1$ (off) & $1.0$ & $16$ \\
                          & \gm{} & $1.0$ & $1.00$ & $-1$ (off) & $1.0$ & $16$ \\
                          & \qw{} & $1.0$ & $1.00$ & $-1$ (off) & $1.0$ & $16$ \\
    \midrule
    \texttt{hf\_pipeline} & \lm{} & $0.6$ & $0.90$ & $50$ & $1.0$ & $20$ \\
                          & \gm{} & $1.0$ & $0.95$ & $64$ & $1.0$ & $20$ \\
                          & \qw{} & $0.7$ & $0.80$ & $20$ & $1.1$ & $20$ \\
    \midrule
    \texttt{hf\_raw}      & \lm{} & $0.6$ & $0.90$ & $50$ & $1.0$ & $20$ \\
                          & \gm{} & $1.0$ & $0.95$ & $64$ & $1.0$ & $20$ \\
                          & \qw{} & $0.7$ & $0.80$ & $20$ & $1.1$ & $20$ \\
    \midrule
    \texttt{langchain\_hf} & \lm{} & $0.6$ & $0.90$ & $50$ & $1.0$ & $20$ \\
                           & \gm{} & $1.0$ & $0.95$ & $64$ & $1.0$ & $20$ \\
                           & \qw{} & $0.7$ & $0.80$ & $20$ & $1.1$ & $20$ \\
    \bottomrule
  \end{tabular}
\end{table*}

\section{Default generation parameters}

Table~\ref{tab:gen_defaults} shows the default parameters of the models for each backend. Each cell is the effective generation configuration a backend applies when the harness supplies no parameter, read directly from the running stack. For the HuggingFace-family backends from each model's \texttt{model.generation\_config} (\texttt{transformers}~$4.57$), for vLLM~$0.8.5$ from \texttt{SamplingParams()}, and for Ollama~$0.31.1$ from \texttt{ollama show -{}-parameters} on the exact fp16 tags we serve. The three HuggingFace backends dispatch to the same \texttt{model.generate()} and therefore share one configuration, which is the model's own tuned one; vLLM ignores the model's \texttt{generation\_config} and samples from a fixed \texttt{temperature}$=1.0$; Ollama reads the model's Modelfile and otherwise falls back to its runner defaults. Values not set by the model or Modelfile inherit the framework fallback (\texttt{transformers}: \texttt{top\_k}$=50$, \texttt{rep.\ pen.}$=1.0$, \texttt{max\_length}$=20$; Ollama: \texttt{temp.}$=0.8$, \texttt{top\_p}$=0.9$, \texttt{top\_k}$=40$, \texttt{rep.\ pen.}$=1.1$). All configurations sample (\texttt{do\_sample=true} in every model's config; vLLM and Ollama always sample).

\section{Per-Model Signed Divergence from \protect\texttt{hf\_raw}}
\label{app:divergence}

\begin{figure*}[htbp]
  \centering
  \includegraphics[width=\linewidth]{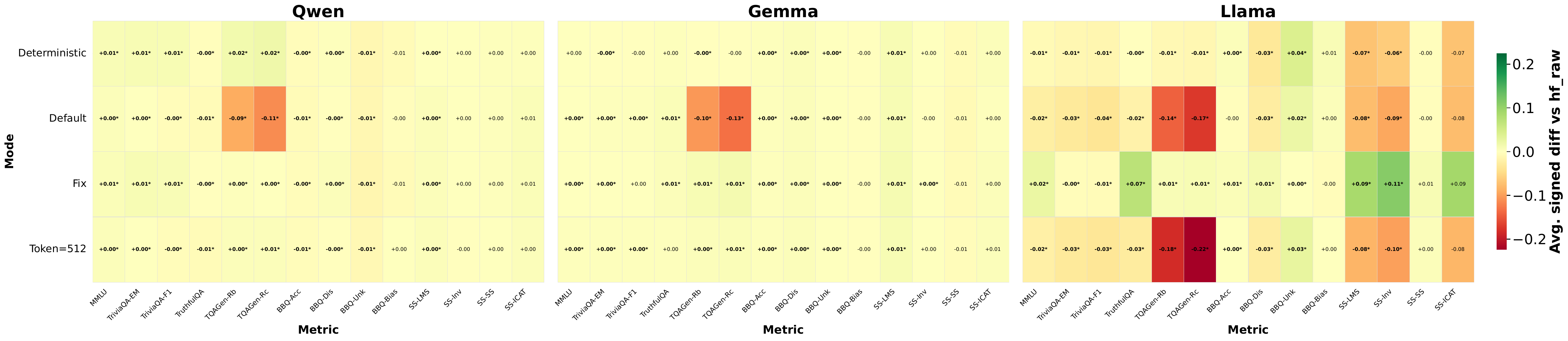}
  \caption{Average divergence from \texttt{hf\_raw} on the calibrated primary metric of each benchmark, for each generation mode. Note that contrary to main body of paper here we report the $\Delta$ instead of $|Delta|$ to capture negative and positive influences. $*$ = at least one backend BH-adjusted significant ($p < 0.05$).}
  \label{fig:app_divergence}
\end{figure*}

Figure~\ref{fig:app_divergence} reveals the \emph{direction} of backend effects that an absolute-value (unsigned) view conceals. For \lm, TruthfulQA cells show mixed positive and negative divergence (Ollama above, vLLM below \texttt{hf\_raw}), explaining why the absolute divergence is large while individual backends go in opposite directions. For Gemma and Qwen, the vast majority of cells are near-zero and non-significant.

\section{Per-Backend Signed Divergence by Task}
\label{app:per_model_div}

\begin{figure*}[htbp]
  \centering
  \includegraphics[width=0.325\linewidth]{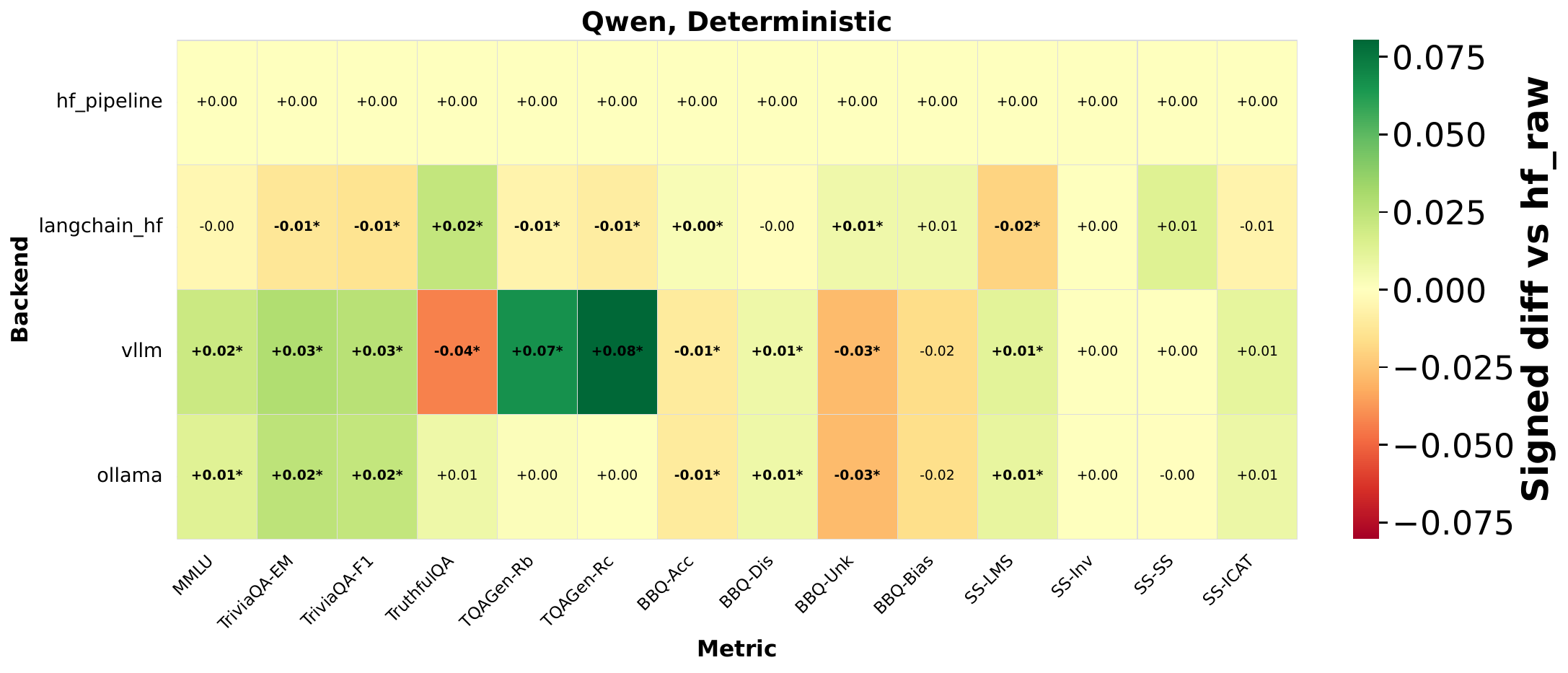}\hfill
  \includegraphics[width=0.325\linewidth]{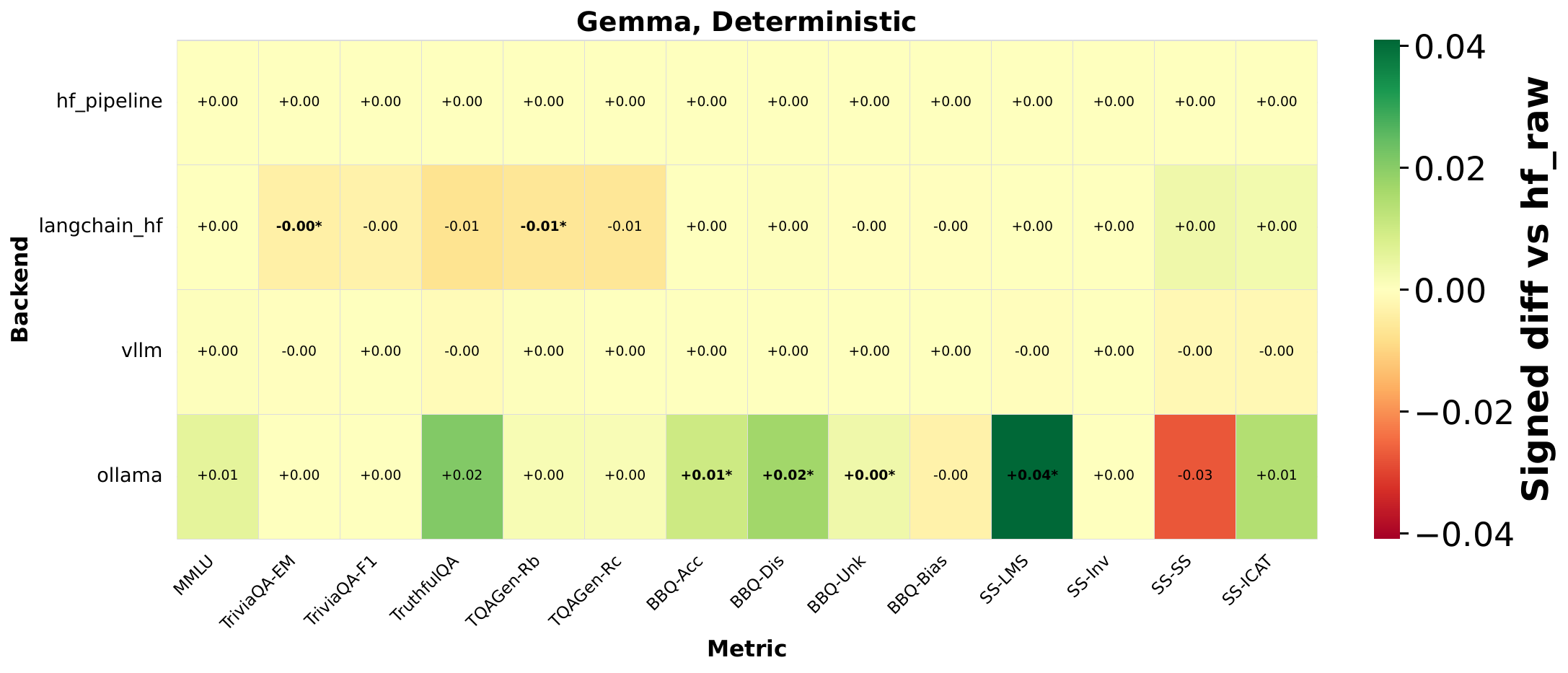}\hfill
  \includegraphics[width=0.325\linewidth]{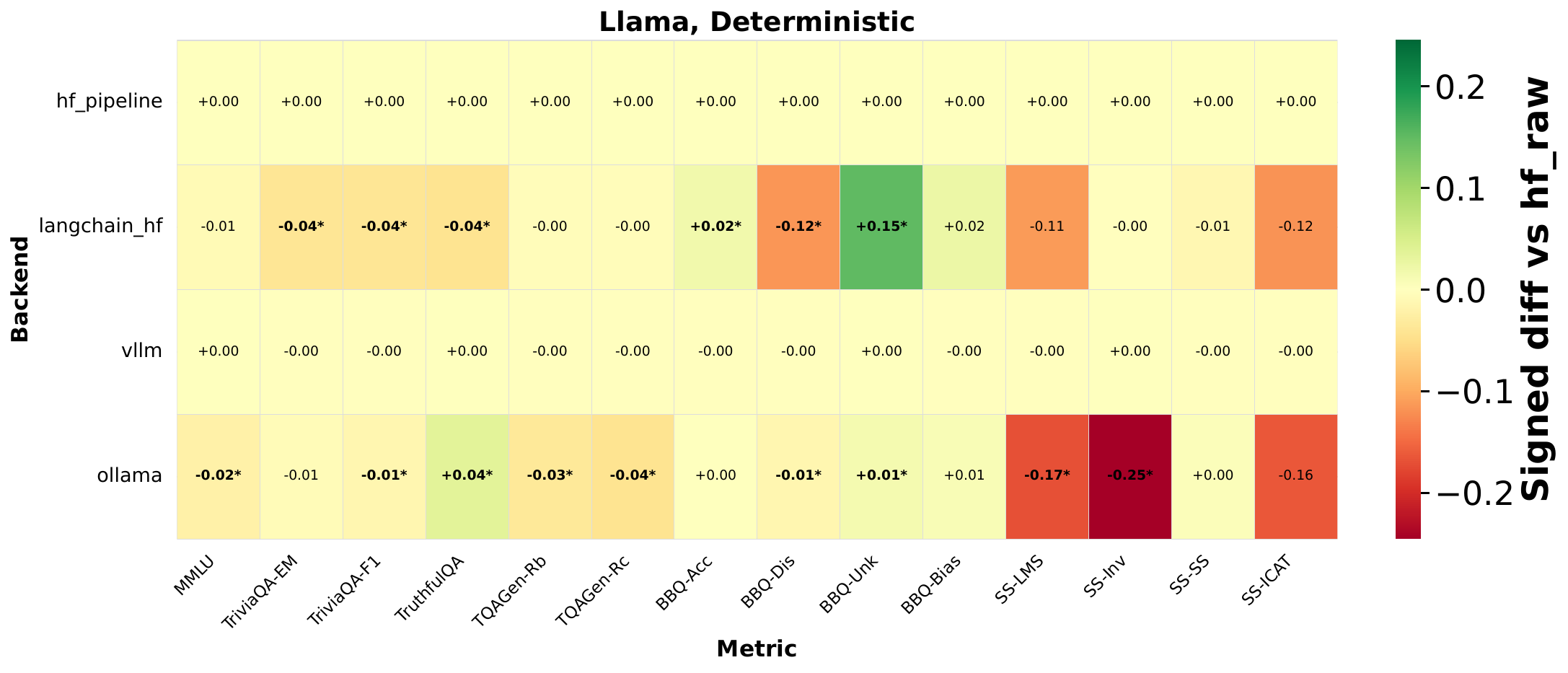}
  \caption{Signed divergence from \texttt{hf\_raw} (green = backend higher, red = lower) per backend (rows) and task (columns), one panel per model family in \textbf{Deterministic} mode. $*$ marks a task with at least one BH-significant metric.}
  \label{fig:permod_det}
\end{figure*}

\begin{figure*}[htbp]
  \centering
  \includegraphics[width=0.325\linewidth]{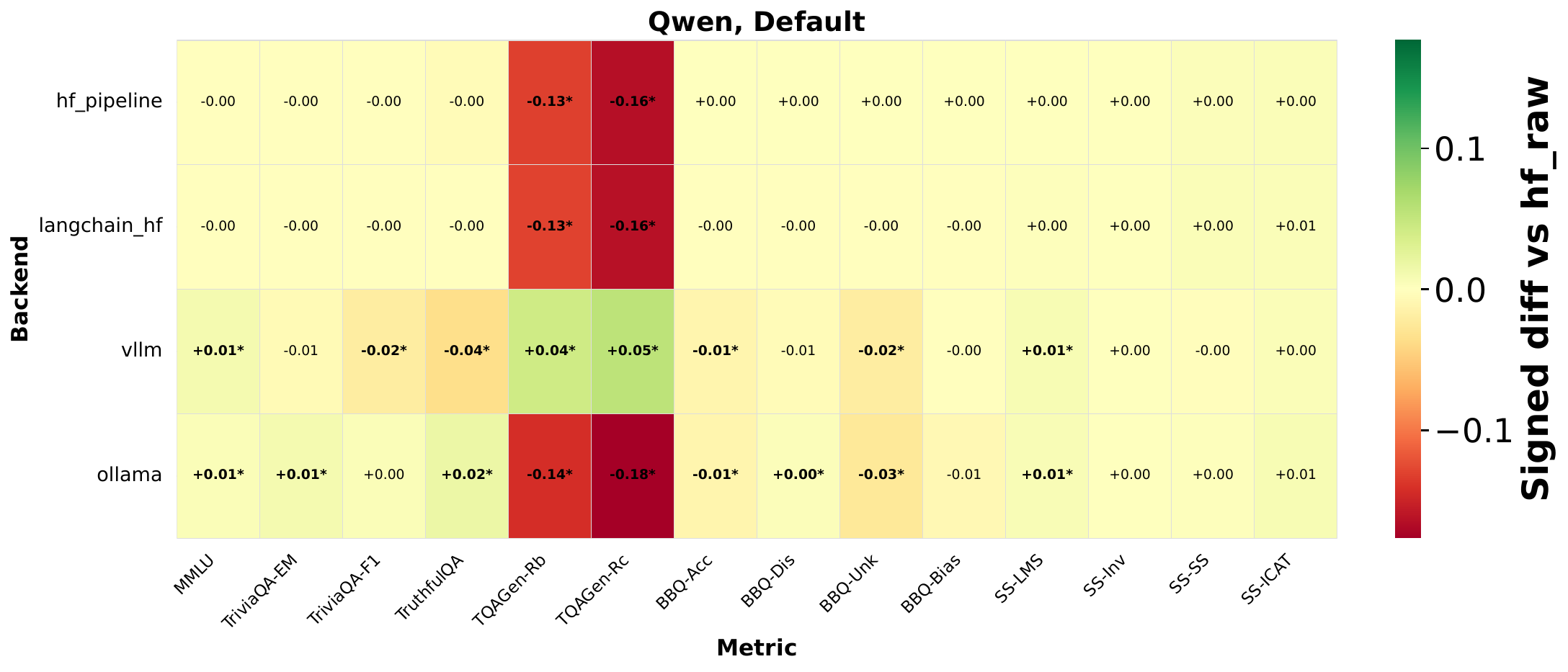}\hfill
  \includegraphics[width=0.325\linewidth]{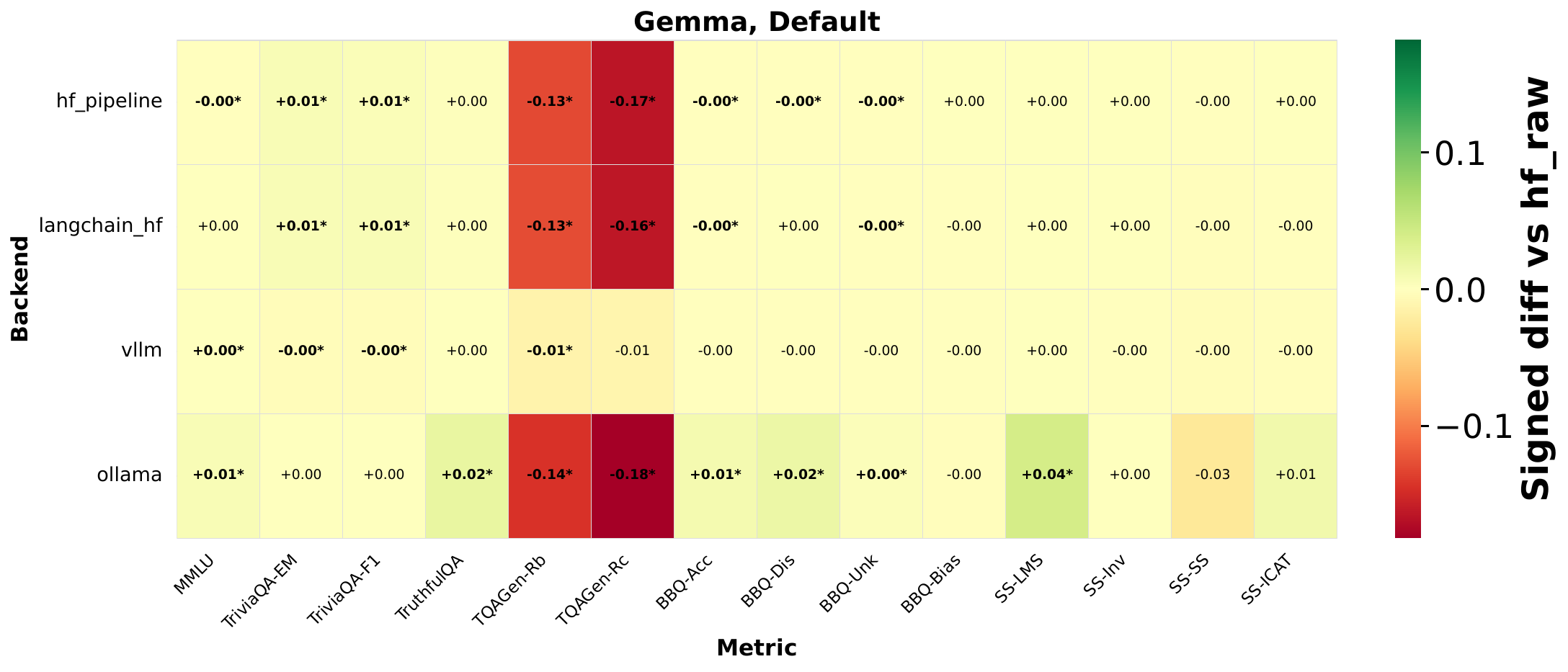}\hfill
  \includegraphics[width=0.325\linewidth]{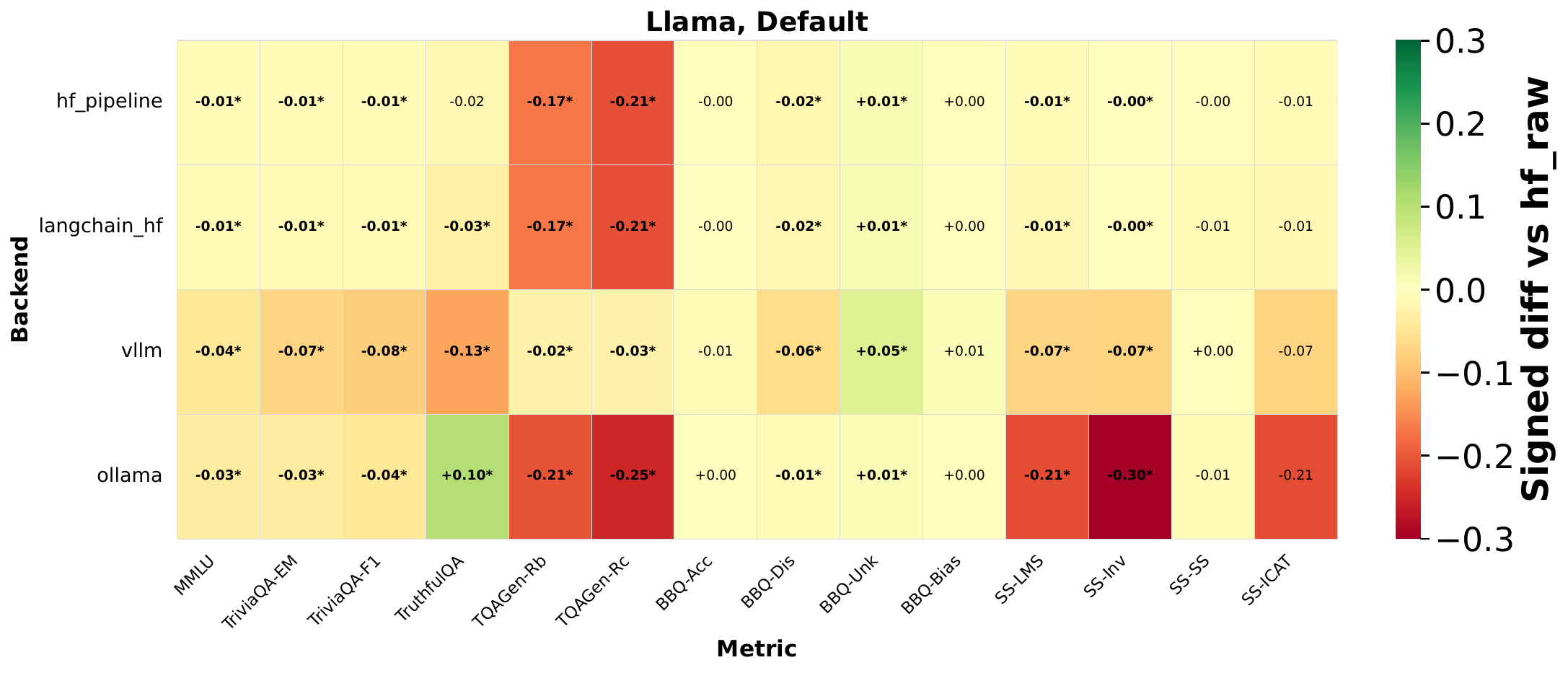}
  \caption{Signed divergence from \texttt{hf\_raw} per backend (rows) and task (columns), one panel per model family in \textbf{Default} mode. Divergences are larger than under Deterministic decoding, especially for \texttt{vllm} and \texttt{ollama} on Llama's factual tasks.}
  \label{fig:permod_def}
\end{figure*}

Figures~\ref{fig:permod_det} and~\ref{fig:permod_def} give the most granular view behind RQ1 and RQ2. For each model, and for the Deterministic and Default modes, the signed difference of every non-baseline backend from \texttt{hf\_raw} on each task's primary metric, with $*$ marking tasks on which at least one metric is BH-significant. Under Deterministic decoding (Figure~\ref{fig:permod_det}) \texttt{hf\_pipeline} is uniformly near zero, Gemma's rows are pale throughout, and Llama shows the strongest structural divergence on the TruthfulQA family. Under Default decoding (Figure~\ref{fig:permod_def}) the cells deepen most sharply for \texttt{vllm} and \texttt{ollama} on Llama's factual tasks reflecting the framework-default sampling gaps and added sampling noise discussed in Section~\ref{sec:rq3}.

\section{Inter-Backend Agreement Analysis}
\label{app:agreement}

\begin{figure*}[htbp]
  \centering
  \includegraphics[width=\linewidth]{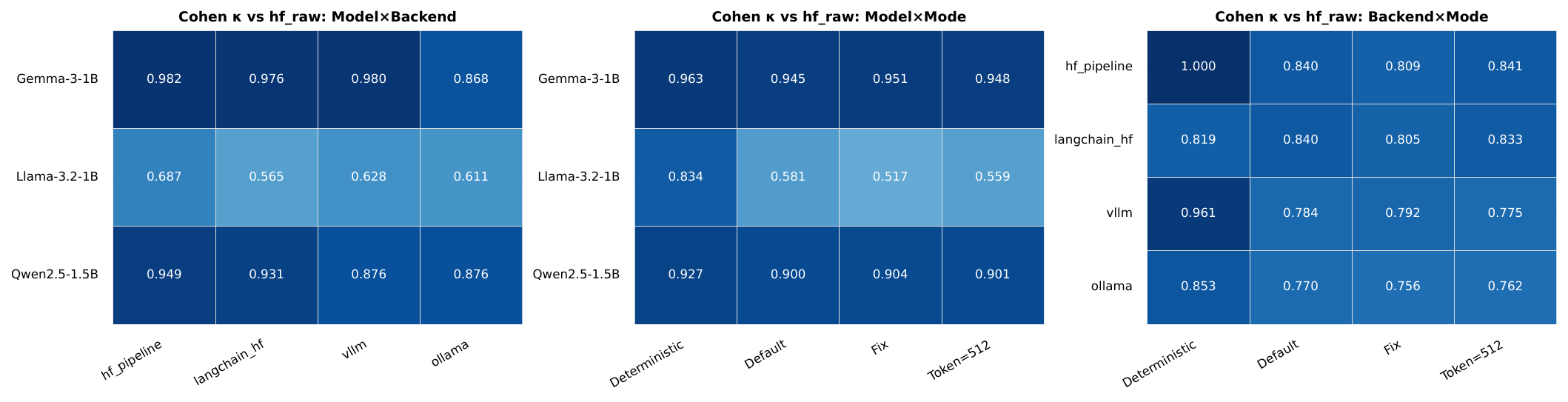}
  \caption{Cohen's $\kappa$ between each non-baseline backend and \texttt{hf\_raw} for correct/wrong answer. Values are averaged over all conditions not held constant.}
  \label{fig:app_agreement}
\end{figure*}

To complement aggregate mean-divergence metrics, we assess whether backends agree on \emph{which questions} a model \textit{answers correctly}. For each (model, mode, dataset) we compute whether models agree on correctness of the answer as a binary measure, correct/wrong and calculate \textbf{Fleiss~$\kappa$} across all five backends jointly, and \textbf{Cohen's~$\kappa$} for each non-baseline backend paired against \texttt{hf\_raw}. Figure~\ref{fig:app_agreement} shows compound Cohen~$\kappa$ views. The Backend $\times$ mode panel (right) confirms the sensitivity ordering: \texttt{hf\_pipeline} achieves perfect agreement with \texttt{hf\_raw} across all models ($\kappa 1.0$), while \texttt{langchain} shows the widest variation ($0.81$) under deterministic follows ollama ($0.85$). How ever these results are model dependent as \lm is the most influential factor in all of these variations where for Langchain the value drops to $0.56$ while for \gm and \qw these values stays at $0.97$ and $0.93$ averaging over all variations respectively.

These $\kappa$ values quantify what the mean-divergence heatmap implies: backends do not merely shift scores uniformly; they change the identity of questions answered correctly, and the degree of change is strongly model-dependent. The severity of this per-question rearrangement is shown in the main text (Figure~\ref{fig:delta_disagree}, Section~\ref{sec:rq2}).

\section{Error-Set Overlap (Jaccard Similarity)}
\label{app:jaccard}

\begin{figure*}[htbp]
  \centering
  \includegraphics[width=\linewidth]{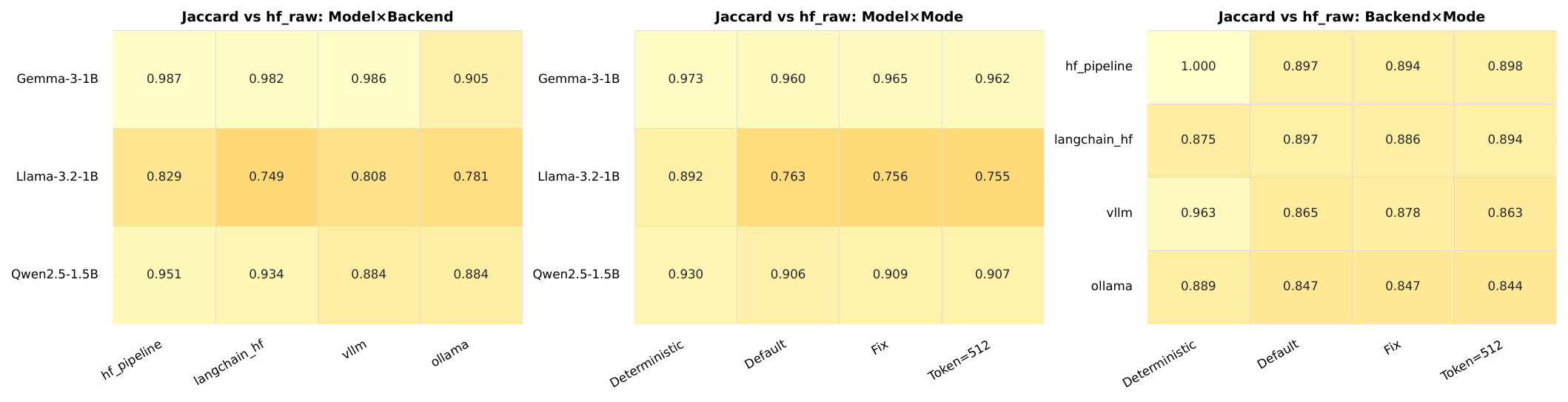}
  \caption{Jaccard similarity of the error sets (incorrectly answered questions) between each non-baseline backend and \texttt{hf\_raw}, shown as compound heatmaps: model $\times$ backend (left), model $\times$ mode (centre), backend $\times$ mode (right). Higher Jaccard = the same questions fail regardless of backend.}
  \label{fig:app_jaccard}
\end{figure*}

Jaccard similarity of error sets measures whether backends fail on the \emph{same} questions as \texttt{hf\_raw}, independently of how many questions fail overall. A Jaccard of 1 means the two backends fail on exactly the same questions while 0 means completely disjoint failure sets.

Figure~\ref{fig:app_jaccard} shows the model $\times$ backend compound view. For \gm, all non-baseline backends maintain Jaccard $\geq 0.93$ against \texttt{hf\_raw}, confirming that Gemma's errors are benchmark-specific and backend-independent. For \lm, mean Jaccard vs.\ \texttt{hf\_raw} drops to $0.792$. Backends are not merely shifting Llama's scores they are redirecting success and failure to different questions, a qualitatively different failure mode. \texttt{hf\_pipeline} again achieves the highest Jaccard across all models, while \texttt{ollama} shows the lowest for \lm{}. The backend $\times$ mode panel shows that stochastic modes reduce Jaccard uniformly (more noise introduces more error-set divergence), but the model-dependent ordering persists in every mode.

\section{Backend Clustering in Score Space}
\label{app:clustering}

\begin{figure*}[htbp]
  \centering
  \includegraphics[width=\linewidth]{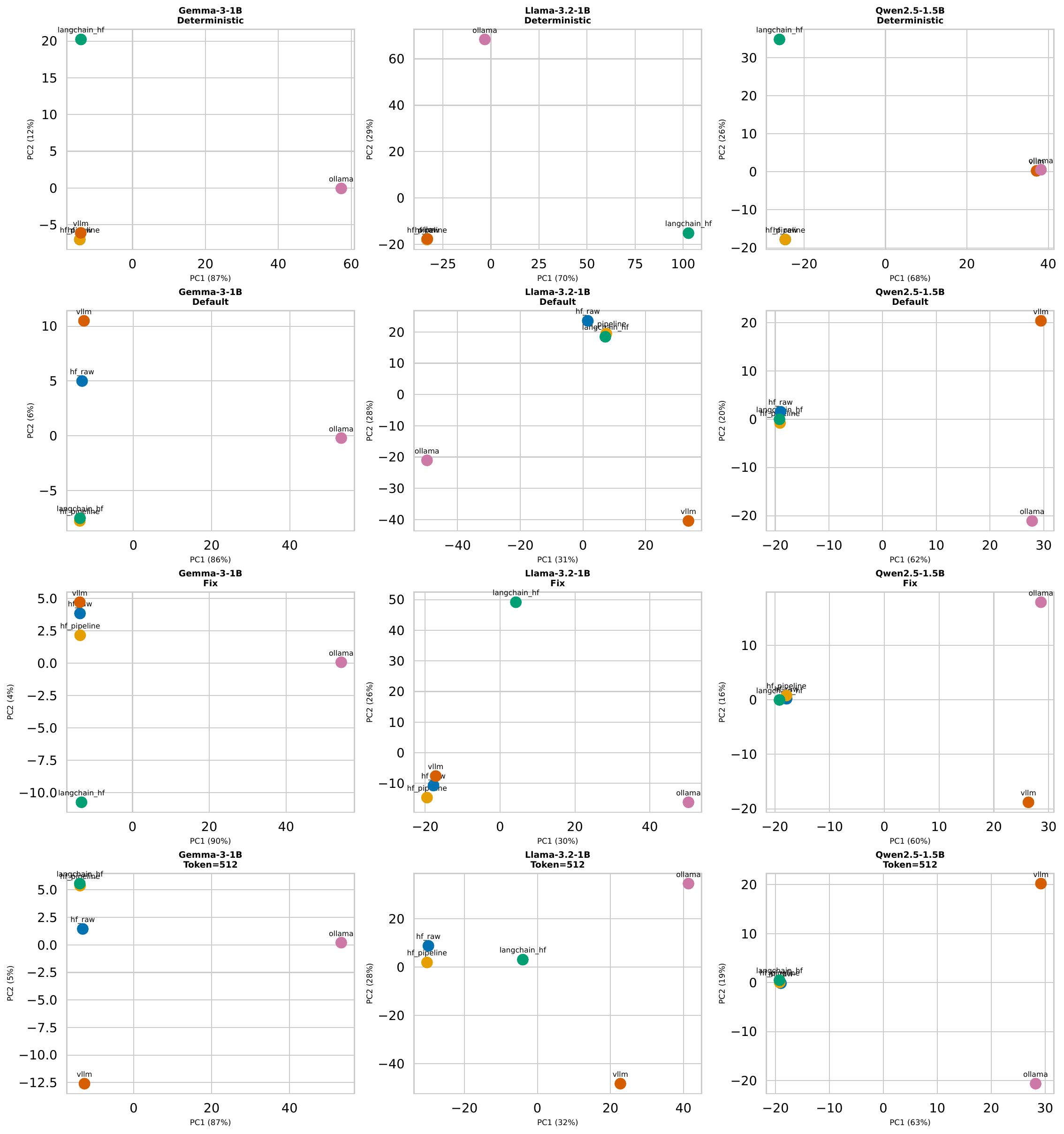}
  \caption{PCA of per-question calibrated score vectors (one vector per backend per model $\times$ mode panel). Each point represents one backend projected onto the first two principal components of the question-score matrix. Backends closer together produce more similar per-question behaviour. Variance explained by PC1/PC2 shown on each axis.}
  \label{fig:app_pca}
\end{figure*}

To understand the geometric relationship between backends in per-question score space, we represent each backend as a vector of calibrated scores over all questions and apply PCA to reduce to two dimensions.

Figure~\ref{fig:app_pca} shows one panel per (model, mode) combination. For \gm{}, all five backends cluster within a compact region in PC1--PC2 space regardless of mode, confirming that they are producing nearly identical per-question responses. For \lm{}, \texttt{ollama} separates along PC1 from the four HuggingFace-ecosystem backends in every mode; under stochastic modes, \texttt{vllm} also separates along PC2. For \qw{}, the cluster is more diffuse than Gemma's but tighter than Llama's.

\section{Response Length as a Potential Confound}
\label{app:length}

\begin{figure*}[t]
  \centering
  \includegraphics[width=\linewidth]{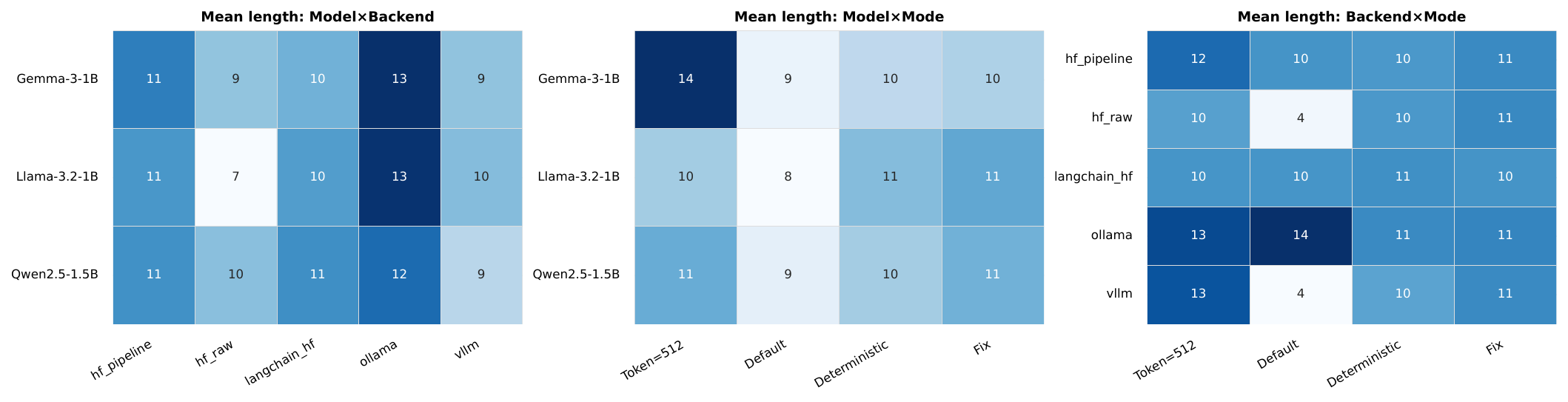}
  \caption{Mean response length (characters) as compound heatmaps: model $\times$ backend (left), model $\times$ mode (centre), backend $\times$ mode (right).}
  \label{fig:app_length}
\end{figure*}

For generative tasks (TriviaQA, TruthfulQA-Gen), response length is a potential confound: a backend that produces longer responses may achieve higher ROUGE-L scores simply due to increased token overlap, not due to better truthfulness or factual recall.

Figure~\ref{fig:app_length} shows mean response length across all tasks. The backend $\times$ mode panel reveals that Default mode produces the highest length variance across backends (since each backend uses its own default \texttt{max\_new\_tokens}), while Fix and Deterministic modes are length-matched. The model $\times$ backend panel shows that \texttt{langchain\_hf} and \texttt{vllm} tend to produce longer responses than \texttt{hf\_raw} for \lm{} and \qw{} under unconstrained modes, consistent with the positive ROUGE-L divergence seen for those backends on TruthfulQA-Gen in Default mode.

A Spearman correlation between response length and calibrated score on generative tasks reveals a positive but moderate relationship for TriviaQA and TruthfulQA-Gen: longer responses are associated with higher scores, but the relationship is driven by a few failure cases (very short or truncated outputs) rather than a monotonic length-quality trade-off. Researchers using ROUGE-L as a proxy for generation quality should therefore ensure that \texttt{max\_new\_tokens} is matched across all comparison frameworks before drawing cross-backend conclusions.

\section{Effect Size}
\label{sec:effect_size}
As additional assurance that the significance tests are not merely minor changes inflated by the large scale of the datasets, we provide an effect-size analysis per model, mode, and backend. Each cell is the mean absolute matched-pairs rank-biserial correlation $|r|$ between a backend and \texttt{hf\_raw}, averaged over the eleven per-item metrics, for each model, backend, and generation mode. Rank-biserial is the paired effect size of the per-item test (McNemar for binary metrics, Wilcoxon signed-rank for continuous ones) where $|r|<0.1$ negligible, $0.1$--$0.3$ small, $0.3$--$0.5$ medium, $>0.5$ large. \texttt{hf\_pipeline} shares \texttt{hf\_raw}'s in-process engine and is an exact no-op under Deterministic decoding. Because $|r|$ is computed only over items on which a backend disagrees with \texttt{hf\_raw}, it captures the \emph{direction} of the change rather than its volume, and should be read together with the divergence magnitudes in Table~\ref{tab:det_div}

\section{Usage of LLM}

All the contents of the paper are written and investigated by the authors. We have used Claude code for visualization of the analysis and help as writing tool.

\begin{table}[htbp]
  \centering
  \small
  \resizebox{\columnwidth}{!}{%
  \begin{tabular}{lcccc}
    \toprule
    Backend & Det. & Fix & Token=256 & Default \\
    \midrule
    \multicolumn{5}{l}{\textbf{\lm}} \\
    \texttt{hf\_pipeline} & 0.000 & 0.021 & 0.394$^{10}$ & 0.380$^{9}$ \\
    \texttt{langchain\_hf} & 0.388$^{6}$ & 0.338$^{8}$ & 0.422$^{10}$ & 0.357$^{10}$ \\
    \texttt{vllm} & 0.112 & 0.051$^{4}$ & 0.517$^{10}$ & 0.448$^{10}$ \\
    \texttt{ollama} & 0.338$^{9}$ & 0.190$^{9}$ & 0.419$^{10}$ & 0.478$^{10}$ \\
    \midrule
    \multicolumn{5}{l}{\textbf{\gm}} \\
    \texttt{hf\_pipeline} & 0.000 & 0.082 & 0.118$^{5}$ & 0.344$^{8}$ \\
    \texttt{langchain\_hf} & 0.149$^{2}$ & 0.223$^{4}$ & 0.219$^{5}$ & 0.323$^{6}$ \\
    \texttt{vllm} & 0.124 & 0.072 & 0.145$^{1}$ & 0.131$^{4}$ \\
    \texttt{ollama} & 0.150$^{4}$ & 0.324$^{10}$ & 0.219$^{5}$ & 0.384$^{8}$ \\
    \midrule
    \multicolumn{5}{l}{\textbf{\qw}} \\
    \texttt{hf\_pipeline} & 0.036 & 0.021$^{2}$ & 0.061$^{2}$ & 0.196$^{2}$ \\
    \texttt{langchain\_hf} & 0.270$^{8}$ & 0.048$^{4}$ & 0.028 & 0.193$^{2}$ \\
    \texttt{vllm} & 0.383$^{10}$ & 0.203$^{8}$ & 0.277$^{8}$ & 0.201$^{8}$ \\
    \texttt{ollama} & 0.218$^{7}$ & 0.157$^{8}$ & 0.167$^{8}$ & 0.299$^{9}$ \\
    \bottomrule
  \end{tabular}
  }
  \caption{Effect size of models for each backend and generation mode. A superscript gives the number of the six metrics that are BH-significant ($p<0.05$) in that cell.}
  \label{tab:effect_size_modes}
\end{table}



\end{document}